\overfullrule=0pt
\input harvmac
\def\a{{\alpha}}

\def\l{{\lambda}}
\def\lb{{\overline\lambda}}
\def\b{{\beta}}

\def\g{{\gamma}}

\def\e{{\epsilon}}

\def\half{{1\over 2}}
\def\p{{\partial}}

\def\t{{\theta}}
\def\tb{{\overline\theta}}
\def\bar{\overline}

\Title{\vbox{\hbox{IFT-P.031/2006 }}}
{\vbox{
\centerline{\bf New Higher-Derivative $R^4$ Theorems}}}
\bigskip\centerline{Nathan Berkovits\foot{e-mail: nberkovi@ift.unesp.br}}
\bigskip
\centerline{\it Instituto de F\'\i sica Te\'orica, State University of
S\~ao Paulo}
\centerline{\it Rua Pamplona 145, 01405-900, S\~ao Paulo, SP, Brasil}

\vskip .3in
The non-minimal pure spinor formalism for the superstring is used
to prove two new multiloop theorems which are related to recent
higher-derivative $R^4$ conjectures of Green, Russo and Vanhove. The first 
theorem states that when $0<n<12$, $\p^n R^4$ terms in the Type II
effective action do not receive perturbative contributions above
$n/2$ loops. The second theorem states that when $n\leq 8$, 
perturbative contributions to
$\p^n R^4$ terms in the IIA and IIB effective actions coincide.

\vskip .3in

\Date {August 2006}

\newsec{Introduction}

Computation of superstring multiloop amplitudes is useful for verifying
perturbative finiteness and for testing non-perturbative duality
conjectures. Although there are several prescriptions available for
computing multiloop amplitudes, the most efficient prescription is
based on the pure spinor formalism which is manifestly super-Poincar\'e
covariant and which involves a bosonic ghost $\l^\a$ satisfying the pure
spinor constraint $\l\g^m\l=0$ \ref\pureone{N. Berkovits,
{\it Super-Poincar\'e covariant quantization of the superstring},
JHEP 0004 (2000) 018, hep-th/0001035.}.

Using the minimal version of the pure spinor formalism, a multiloop
prescription was defined in \ref\multione{N. Berkovits, {\it Multiloop 
amplitudes and vanishing theorems using the pure spinor formalism for
the superstring}, JHEP 0409 (2004) 047, hep-th/0406055.}
and used to prove certain vanishing theorems. One theorem
stated that massless $N$-point multiloop amplitudes are vanishing when
$N\leq 3$, which is related to perturbative finiteness of the superstring
\ref\mart{E.J. Martinec, {\it Nonrenormalization theorems and fermionic
string finiteness}, Phys. Lett. B171 (1986) 189\semi
S. Mandelstam, {\it The n loop string amplitude: Explicit formulas,
finiteness and absence of ambiguities.}, Phys. Lett. B277 (1992) 82.}.
Another theorem stated that $R^4$ terms in the Type II
effective action do not receive perturbative contributions above one-loop,
which is related to S-duality of the Type IIB superstring 
\ref\gut{M.B. Green and M. Gutperle, {\it Effects of D instantons},
Nucl. Phys. B498 (1997) 195, hep-th/9701093.}\ref\gvone{M.B. Green
and P. Vanhove, {\it D-instantons, strings and M-theory}, Phys. Lett. 
B408 (1997) 122, hep-th/9704145.}. In later papers,
four-point one and two-loop amplitudes were computed using the minimal
formalism and were shown to coincide with the RNS result \ref\nbtwo{
N. Berkovits, {\it Super-Poincar\'e covariant two-loop superstring amplitudes},
JHEP 0601 (2006) 005, hep-th/0503197\semi
N. Berkovits and C.R. Mafra, {\it Equivalence of two-loop superstring
amplitudes in the pure spinor and RNS formalisms}, Phys. Rev. Lett. 96 
(2006) 011602, hep-th/0509234\semi C.R. Mafra, {\it Four-point
one-loop amplitude computation in the pure spinor formalism}, JHEP
0601 (2006) 075, hep-th/0512052.}.

Recently, a new multiloop prescription was proposed using a non-minimal
version of the pure spinor formalism which involves both the 
pure spinor $\l^\a$ and its complex conjugate $\lb_\a$ \ref\puretopo
{N. Berkovits, {\it Pure spinor formalism as an N=2 topological
string}, JHEP 0510 (2005) 089, hep-th/0509120.}.
This non-minimal prescription for multiloop amplitudes
has several advantages over the minimal 
prescription. Firstly, unlike in the minimal formalism, the non-minimal
formalism allows the construction of a composite $b$ ghost satisfying
$\{Q, b\}=T$. Secondly, the non-minimal formalism can be interpreted as
a critical topological string, so the non-minimal amplitude prescription is the
same as in bosonic string theory. Thirdly, there is no need for 
picture-changing operators in the non-minimal prescription, which were
inconvenient in the minimal prescription since they broke 
manifest Lorentz covariance at intermediate
stages in the computation.

The only difficulty in the non-minimal prescription is regularizing
the functional integral over the pure spinor ghosts when $\l\to 0$, 
which will be discussed in an upcoming 
paper with Nikita Nekrasov \ref\nekra{N. Berkovits and N. Nekrasov,
{\it Multiloop superstring amplitudes from 
non-minimal pure spinor formalism}, hep-th/0609012.}.
Fortunately, this $\l\to 0$ regularization
is unnecessary for proving the multiloop theorems in this paper.
Furthermore, it was recently shown with Carlos
Mafra that various one and two-loop amplitude computations using the
non-minimal prescription correctly reproduce the RNS result \ref\mtwo
{N. Berkovits and C.R. Mafra, {\it Some superstring amplitude computations
with the non-minimal pure spinor formalism}, hep-th/0607187.}. Although
it should be possible to directly prove the equivalence of the minimal
and non-minimal amplitude prescriptions (perhaps using the Cech description
of Nekrasov \ref\jeru{N. Nekrasov, {\it Lectures at the
23rd Jerusalem Winter School in Theoretical Physics, January 2-5, 2006},
http://www.as.huji.ac.il/schools/phy23/media2.shtml.}), 
this has not yet been done.

In this paper, the non-minimal prescription will be used to prove 
two new multiloop theorems which are related
to recent higher-derivative
$R^4$ conjectures of Green, Russo and Vanhove based on
duality symmetries \ref\grv{M.B. Green, J. Russo and P. Vanhove, {\it to 
appear}\semi M.B. Green, {\it talk at Strings 2006}.}
\ref\greenf{M.B. Green and P. Vanhove, {\it Duality and higher derivative
terms in M theory}, JHEP 0601 (2006) 093, hep-th/0510027.}
\ref\greeng{M.B. Green and P. Vanhove, {\it The low-energy expansion of
the one loop type II superstring amplitude}, Phys. Rev. D61 (2000)
104011, hep-th/9910056.}
\ref\greenh{M.B. Green, H. Kwon and P. Vanhove, {\it Two loops in eleven
dimensions}, Phys. Rev. D61 (2000) 104010, hep-th/9910055.}
\ref\greenj{M.B. Green, M. Gutperle and P. Vanhove, {\it One loop in eleven
dimensions}, Phys. Lett. B409 (1997) 177, hep-th/9706175.}
\ref\dixon{Z. Bern, L. Dixon, D. Dunbar. M. Perelstein and J.S. Rozowsky,
{\it On the relationship between Yang-Mills theory and gravity and its
implication for ultraviolet divergences}, Nucl. Phys. B530 (1998) 401,
hep-th/9802162.}.
The first new multiloop theorem states that when $0<n<12$, $\p^n R^4$
terms in the Type II effective action do not receive genus
$g$ contributions for $g>n/2$. 
The restriction that $n<12$ is related to the fact that
$\p^{12} R^4$ 
can be written as a superspace integral over 32 $\t$'s. 
The second new multiloop theorem states that when $n\leq 8$,
perturbative contributions to $\p^n R^4$ terms in the IIA and IIB effective
action coincide. For $n\leq 4$, this can be shown using the RNS formalism
\greenh,
and for $n=6$, it was recently conjectured by Green and Vanhove
\greenf.

In section 2, the amplitude prescription using the non-minimal
formalism will be reviewed and, in section 3, the
two new multiloop theorems will be proven.

\newsec{Review of Non-Minimal Amplitude Prescription}

\subsec{Amplitude prescription}

Using the non-minimal pure spinor formalism, the 
$N$-point $g$-loop 
ampitude prescription is \puretopo
\eqn\corr{{\cal A} = \int d^{6g-6}\tau \langle
|\prod_{j=1}^{3g-3}(\int dy_j \mu_j(y_j) b(y_j))|^2 \prod_{r=1}^N 
\int d^2 z_r U_r(z_r)  ~|{\cal N}|^2\rangle}
where $\int d^{6g-6}\tau 
\langle
|\prod_{j=1}^{3g-3}(\int dy_j \mu_j(y_j) b (y_j))|^2 
\prod_{r=1}^N 
\int d^2 z_r U_r(z_r) \rangle$ is the usual amplitude prescription of
bosonic string theory, $\tau_j$ and $\mu_j$ are the
Teichmuller parameters and corresponding Beltrami differentials,
$b$ is a composite operator
satisfying $\{Q, b\}=T$, and ${\cal N}= e^{\{Q, \chi\}}$ is
a BRST-invariant operator which regularizes the $0/0$ coming from
integration over the worldsheet zero modes. 

The left-moving
worldsheet variables in the non-minimal pure spinor formalism
include the Green-Schwarz-Siegel variables $(x^m, \t^\a, d_\a)$,
the pure spinor ghosts $(\l^\a, w_\a)$, and the non-minimal variables
$(\lb_\a, \bar w^\a, r_\a, s^\a)$. In terms of these variables, the composite
$b$ operator is 
\eqn\defbg{b =
s^\a\p\lb_\a  + {{\lb_\a (2
\Pi^m (\g_m d)^\a-  N_{mn}(\g^{mn}\p\t)^\a
- J \p\t^\a -{1\over 4} \p^2\t^\a)}\over{4(\lb\l)}}  }
$$+ {{(\lb\g^{mnp} r)(d\g_{mnp} d +24 N_{mn}\Pi_p)}\over{192(\lb\l)^2}}
-
{{(r\g_{mnp} r)(\lb\g^m d)N^{np}}\over{16(\lb\l)^3}} +
{{(r\g_{mnp} r)(\lb\g^{pqr} r) N^{mn} N_{qr}}\over{128(\lb\l)^4}} $$
where $\Pi^m = \p x^m + \t\g^m \p\t$ is the supersymmetric momentum,
and $N_{mn}=\half w\g_{mn}\lambda$ and $J=\l w$ are the
Lorentz and ghost currents for the pure spinors.
Although the expression of \defbg\ is complicated, it will turn out
that only two terms in \defbg\ are relevant for proving the 
multiloop theorems, namely the terms $ {{\Pi^m(\lb\g_m d_\a)}\over{2(\l\lb)}}$
and  
${{(\lb\g^{mnp} r)(d\g_{mnp} d)}\over{192(\lb\l)^2}}$.

The
zero-mode regulator
${\cal N}$ in \corr\ can be defined as
\eqn\calNloop{{\cal N} =
\exp {\{Q, \chi\}} }
$$=\exp [-\l^\a\lb_\a -r_\a \t^\a
-
\half N_{mn}^I \bar N^{mn I} - J^I \bar J^I -
{1\over 4}(s^I\g^{mn}\lb)(d^I\g_{mn}\l)
-(\lb s^I) (\l d^I)~], $$
where $s^{\a I}= \oint_{a_I} dz s^\a$, 
$d_{\a}^I= \oint_{a_I} dz d_\a$, 
$N_{mn}^I=\oint_{a_I}dz N_{mn}$, $J^I=\oint_{a_I}dz J$ are the
zero modes 
obtained by integrating these fields around the $I^{th}$
$a$-cycle, and
\eqn\chinew{\chi 
= -\lb_\a \t^\a - \half N_{mn}^I (s^I\g^{mn} \lb)  - J^I 
(\lb s^I).}
When the bosonic zero modes of $\l^\a$, $N_{mn}^I$ or $J^I$ go to infinity, 
the functional integral over these zero modes is well-defined
because of the exponential cutoff in $\cal N$. 

However, when $\l^\a\to 0$,
the poles in \defbg\ 
make the functional integral over $\l^\a$ 
ill-defined if the sum of the degree of the poles is greater than or equal to
11. If the contributions from the $b$ ghosts diverge as fast as $(\l\lb)^{-11}$,
the measure factor $\int d^{11}\l d^{11}\lb $ does not converge fast
enough to make the functional integral well-defined.
In an upcoming paper with Nekrasov \nekra, it will be shown how
to regularize this $\l\to 0$ divergence for arbitrary multiloop
amplitudes.
However, there are certain amplitudes for which the sum of the degree of the 
poles from the $b$ ghosts
is always less than 11, so one does not need to worry about regularizing the 
$\l\to 0$ divergence. 
For example, since the maximum pole in the $b$ ghost is of degree 3,
there is no $\l\to 0$ divergence when the genus is less than or equal to
two since, for
these amplitudes, there are three
or fewer $b$ ghosts.
As will also be discussed in
\nekra, another type of amplitude for which
there is no $\l\to 0$ divergence is when at least one of the
sixteen $\t^\a$ zero modes comes from $\cal N$. 

\subsec{F-terms}

To show that terms receiving at least one $\t$ zero mode from ${\cal N}$ do
not contain divergences when $\l\to 0$, first note that BRST invariance
implies that the
$e^{-(\l\lb+ r \t)}$ term in ${\cal N}$ can be modified to
$e^{-\rho(\l\lb+ r \t)}$ for any positive constant $\rho$. Because
\eqn\argrho{e^{-\rho(\l\lb+ r \t)} =
e^{\{Q, -\rho \t\lb\}} = 1 + \{Q,\xi_\rho\} }  
for some $\xi_\rho$, BRST-invariant ampitudes are independent
of the value of $\rho$.

Suppose one computes the amplitude $\langle F(\l,\lb) ~{\cal N}\rangle$
where
$F(\l,\lb)$ is some BRST-invariant operator. Then $\rho$-independence 
implies that the
$(-\rho r \t)^n$ terms in
$$e^{-\rho(\l\lb+ r \t)}
= e^{-\rho\l\lb} (1 + \sum_{n=1}^{11} {1\over{n!}}(-\rho r\t)^n)$$
can only contribute to $\langle F(\l,\lb)~ {\cal N}\rangle$
if $\int d^{11}\l d^{11}\lb~ F(\l,\lb) ~e^{-\rho\l\lb}$ has poles in $\rho$.
But this implies that $F(\l,\lb)$ diverges slower than $(\l\lb)^{-11}$
since
\eqn\divergs{\int d^{11}\l d^{11}\lb~ (\l\lb)^{-n} ~ e^{-\rho\l\lb}
\sim \rho^{n-11}.}
So $\t$ zero modes in ${\cal N}$ can only contribute to
$\langle F(\l,\lb)~ {\cal N}\rangle$
if $F(\l,\lb)$ diverges slower than $(\l\lb)^{-11}$, which implies
that 
$\int d^{11}\l d^{11}\lb F(\l,\lb)$ is well-defined near $\l=0$.

Since the $b$ ghost of \defbg\ is spacetime supersymmetric, the sixteen
$\t^\a$ zero modes in the functional integral of \corr\ must come
either from the regulator ${\cal N}$ or from the external vertex
operators $\prod_{r=1}^N U_r(z_r)$. If all 
sixteen $\t$ zero modes
come from the superfields in the vertex operators $U_r$,
the resulting term in the effective action is not a ten-dimensional
F-term since
it can be written as an integral over the maximum number of $\t$'s.
However, if at least one of the $\t$ zero modes come from ${\cal N}$,
the amplitude could contribute to F-terms.
Therefore, the above argument implies that
amplitudes which contribute to ten-dimensional F-terms
do not require regularization when $\l\to 0$.

Note that as in lower dimensions, D=10 F-terms are
defined as manifestly gauge-invariant
terms in the superspace effective action which cannot be written
as integrals over the maximum number of $\t$'s.
Although one does not know how to construct off-shell D=10 superspace
actions,
one can construct higher-derivative
D=10 superspace actions which are functions of on-shell
linearized superfields.

For example, for open superstrings, the massless super-Yang-Mills vertex
operator is 
\eqn\openv{\int dz U = \int dz (A_M(Z) \p Z^M + W^\a(Z) d_\a
+ F_{mn}(Z) N^{mn})}
where $Z^M=(x^m,\t^\a)$, and the gauge-invariant superfield of 
lowest dimension is $W^\a$
whose lowest component is the gluino of dimension $1\over 2$.
Note that $\int dz U$ can be expressed in manifestly gauge-invariant
form by using the normal-coordinate expansion to write
\eqn\norma{\int dz A_M(Z) \p Z^M = \int dz (\half F_{MN}(Z_0) \hat Z^M \p Z^N
+ {1\over 6}\p_P F_{MN}(Z_0) \hat Z^P \hat Z^M \p Z^N + ...)}
where $\hat Z^M = Z^M - Z_0^M$ and $Z_0^M$ is a constant.
Since
N=1 D=10 superspace contains 16 $\t$'s, any term in the superspace
action involving $M$ superfields $W^\a$ which is integrated over
the full superspace has dimension $\geq (M+16)/2.$ Therefore, any
term in the N=1 D=10 superspace action involving $M$ field-strengths
which has
dimension less than $(M+16)/2$ is necessarily an N=1 D=10 F-term.

For closed Type IIB superstrings,
the massless supergravity vertex operator is
\eqn\closedv{\int d^2 z U = \int dz ((G_{MN}(Z)+B_{MN}(Z)) 
\p Z^M \bar\p Z^N + W^{\a\b}(Z) d_\a \bar d_\b + ...)}
where $Z^M=(x^m,\t^\a,\tb^\b)$, and the gauge-invariant superfield of 
lowest dimension is $W^{\a\b}(x,\t,\tb)$ whose lowest
component is the Ramond-Ramond field strength of dimension 1. Note that
the dilaton and axion are gauge-invariant fields of
dimension zero, but they always appear
with derivatives in the massless vertex operator. Since N=2 D=10
superspace contains 32 $\t$'s,
any term in the superspace
action involving $M$ superfields $W^{\a\b}$ which is integrated over
the full superspace has dimension $\geq (M+16).$ Therefore, any
term in the N=2 D=10 superspace action involving $M$ field-strengths
which has
dimension less than $(M+16)$ is necessarily an N=2 D=10 F-term.
For example, since the curvature tensor $R_{mnpq}$ has dimension 2,
the term
\eqn\arg{\int d^{10}x \sqrt{g} \p^L R^M}
in the Type II effective
action is an N=2 D=10 F-term if $L+2M< M+16$, i.e. if $L+M<16$.

\newsec{New Multiloop Theorems}

The multiloop theorems in this paper will be proven by counting
fermionic zero modes in the integrand of \corr. The left-moving fermionic
worldsheet fields in the non-minimal formalism include the 
superspace variable $\t^\a$ and its conjugate momentum $d_\a$,
and the non-minimal variable $r_\a$ and its conjugate momentum $s^\a$.
The non-minimal variable $r_\a$ is constrained to satisfy $\lb\g^m r=0$
where $\lb\g^m\lb=0$, so $r_\a$ has 11 independent components. Since
$\t^\a$ and $r_\a$ are worldsheet scalars, on a genus $g$ surface
$\t^\a$ contains 16 zero modes, $d_\a$ contains $16g$ zero modes,
$r_\a$ contains 11 zero modes, and $s^\a$ contains $11g$ zero modes.
So 
the amplitude of \corr\ vanishes unless all of these fermionic
zero modes are present in the integrand of \corr.

\subsec{Nonrenormalization of $\p^n R^4$}

Using the prescription of \corr,
it will now be proven that perturbative contributions
to $\p^n R^4$ terms vanish above $n/2$ loops. This is proven by showing
that the massless four-point $g$-loop amplitude
at low energies is proportional to 
\eqn\propo{({\p\over{\p\t}}{\p\over{\p\tb}})^{2g+4} W^4 = \p^{2g} R^4 
+ ...}
where $W$ is the Ramond-Ramond field strength of \closedv.
So $\p^n R^4$ terms only get perturbative contributions up to genus $n/2$.
When $g\geq 6$, \propo\ is no longer an F-term, 
so there may be $\l\to 0$ divergences which need to be regularized.
The theorem has therefore only been proven
when $n< 12$. 

To get a non-vanishing four-point $g$-loop amplitude, the integrand of
\corr\ must provide $16g$ $d_\a$ zero modes which can come either from
the four vertex operators of \closedv, from the regulator ${\cal N}$ of 
\calNloop, or from the $3g-3$ $b$ ghosts of \defbg.
The most efficient way to obtain these $16g$ zero modes is if the 
four vertex operators provide the term $(W^{\a\b} d_\a \bar d_\b)^4$,
and the regulator ${\cal N}$ provides the term
$(s d)^{11g}$, where $11g$ is the maximum power since there are only
$11g$ independent $s$ zero modes.

The remaining $5g-4$ $d_\a$ zero modes must come from the $3g-3$
$b$ ghosts, and to minimize the number of $\t^\a$ zero modes coming
from the vertex operators, it will be advantageous to minimize
the number of $r_\a$ zero modes coming from the $b$ ghosts. The ghost
contribution which provides $5g-4$ $d_\a$ zero modes while minimizing
the number of $r_\a$ zero modes is 
\eqn\ghostc{({{\p x^m(\lb\g_m d)}
\over{2(\lb\l)}} )^{g-2}~ 
( {{(\lb\g^{mnp} r)(d\g_{mnp} d)}\over{192(\lb\l)^2}})^{2g-1},}
where $g-2$ $b$ ghosts provide the first term and $2g-1$ $b$
ghosts provide the second term.

The contribution of \ghostc\ provides $2g-1$ of the 11 $r_\a$
zero modes, so the remaining $12-2g$
$r_\a$ zero modes must come from ${\cal N}$
through the term $(r\t)^{12-2g}$. Since this term provides $12-2g$ of
the 16 $\t^\a$ zero modes, the remaining $2g+4$ $\t^\a$ zero modes must
come from the superfields $W^{\a\b}$ in the vertex operators.

So for the amplitude to be non-vanishing, the four external vertex
operators must provide at least $2g+4$ $\t^\a$ zero modes. Therefore,
at low energies, the four-point $g$-loop scattering amplitude is
proportional to \propo\ as claimed. Note that it is assumed that
that there are no inverse factors of momentum coming from the
$\langle \prod_{r=1}^4 e^{ik_r x(z_r)} \rangle$ correlation function which
would decrease the number of derivatives on $R^4$ in \propo.
This assumption is reasonable since the massless three-point
multiloop amplitude vanishes, so one does not expect any poles in
momentum for the four-point multiloop amplitude.

\subsec{Equivalence of IIA and IIB $\p^n R^4$ terms}

It will now be proven that up to four loops, four-point graviton
contributions to F-terms coincide in the IIA and IIB effective actions.
Using the previous theorem that $\p^n R^4$ terms do
not get perturbative contributions above $n/2$ loops,
this implies that perturbative contributions to $\p^n R^4$ terms
coincide in the IIA and IIB effective actions for $n\leq 8$.

To prove this multiloop theorem, similar methods to 
\greenh\ will be used.
IIA and IIB superstrings are related by a parity operation
on the left-moving worldsheet variables which flips the chirality
of the left-moving spacetime spinor.
For graviton scattering amplitudes, this parity operation flips the
sign of terms which involve an $\e_{m_1 ... m_{10}}$
tensor coming from the integration over the left-moving variables.

For four-point graviton amplitudes, the only way to contract the
vector indices on such an $\e_{m_1 ... m_{10}}$ tensor is if there is also an 
$\e_{n_1 ... n_{10}}$
tensor coming from the integration over the right-moving variables.
One can then contract the vector indices of the left-moving $\e$ tensor
either with the indices of the right-moving $\e$ tensor or with
the external momenta and polarizations. 

Since there are three independent momenta, $k^m_r$ for $r=1$ to 3,
and four independent polarizations, $h^{mn}_r$ for $r=1$ to 4, 
the minimum number
of indices which must be contracted between the left and right-moving
$\e$ tensors is three. This can be accomplished using the contraction
\eqn\contr{h_1^{m_1 n_1} h_2^{m_2 n_2} h_3^{m_3 n_3} h_4^{m_4 n_4}
k_1^{m_5} k_1^{n_5} 
k_2^{m_6} k_2^{n_6} 
k_3^{m_7} k_3^{n_7} 
\eta^{m_8 n_8}
\eta^{m_9 n_9}
\eta^{m_{10} n_{10} }
\e_{m_1 ... m_{10}}\e_{n_1 ... n_{10}}.}

To contract indices of the left and right-moving $\e$ tensors,
the correlation function must involve factors of $\p x^m$ and
$\bar\p x^n$  since these are the only left and right-moving fields
which can be contracted. For example, in the RNS formalism, these
factors of $\p x^m$ and $\bar\p x^n$ come from the left and right-moving
picture-changing operators. Since $g$-loop RNS amplitudes involve
$2g-2$ left and right-moving picture-changing operators, and since one
needs at least three $\p x$'s and $\bar\p x$'s to perform the contraction
of \contr, the term in \contr\ is only possible when $g\geq 3$.
So the four-point graviton amplitudes in IIA and IIB superstring theory
have been proven to coincide up to two loops using the RNS formalism
\greenh.

Using the prescription of section 2 for F-term computations,
the three factors of
$\p x^m$ and $\bar\p x^n$ can come 
from the term 
${{\p x^m(\lb\g_m d)}
\over{2(\lb\l)}}$ in the $b$ ghost. By counting $d_\a$ zero modes
as in \ghostc, one finds at genus $g$ that the
maximum number of $\p x$ factors is $g-2$.
So the contraction of \contr\ is only possible when $g\geq 5$,
implying that four-point graviton amplitudes contribute equally
to IIA and IIB F-terms when $g\leq 4$.

\vskip 15pt
{\bf Acknowledgements:} I would like to thank Nikita Nekrasov, Jorge
Russo, Pierre Vanhove, and especially Michael Green
for useful discussions. I would also like to thank
the hospitality of Cambridge University and the 2006 Eurostrings
conference where a preliminary version of this paper was presented, and
CNPq grant 300256/94-9, Pronex
grant 66.2002/1998-9, and FAPESP grant 04/11426-0 for partial financial
support.

\listrefs
\end